\raggedbottom
\def\nex{\par\noindent\hang}
\null

\vskip10mm

\centerline{\bf DWARF SPHEROIDAL SATELLITE GALAXIES}
\vskip 1mm
\centerline{\bf and the}
\vskip 1mm
\centerline{\bf GALACTIC TIDAL FIELD}

\vskip 5mm
\centerline{\bf Pavel Kroupa}
\vskip 2mm
\centerline{Astronomisches Rechen-Institut}
\vskip 2mm
\centerline{M{\"o}nchhofstra{\ss}e~12-14, D-69120~Heidelberg, Germany}
\vskip 2mm
\centerline{e-mail: S48@ix.urz.uni-heidelberg.de}

\vskip 30mm

\centerline{Invited talk held at the}
\vskip 1.5mm 
\centerline{18th Meeting of the Graduierten-Kolleg}
\vskip 1.5mm
\centerline{The Magellanic System and other dwarf galaxies --}
\vskip 1.5mm
\centerline{Investigations of small galaxies}
\vskip 1.5mm
\centerline{November 13/14/15 1996 in Bad Honnef}
\vskip 1.5mm
\centerline{Organised by U.Klein, Bonn}

\vfill\eject

\centerline{\bf DWARF SPHEROIDAL SATELLITE GALAXIES}
\vskip 1mm
\centerline{\bf and the}
\vskip 1mm
\centerline{\bf GALACTIC TIDAL FIELD}
\vskip 5mm
\centerline{\bf Pavel Kroupa}
\vskip 2mm
\centerline{Astronomisches Rechen-Institut, Heidelberg}

\vskip 10mm
\centerline{\bf Abstract}
\vskip 5mm
\hang{
The Milky Way is surrounded by nine or more dwarf-spheroidal (dSph)
satellite galaxies that appear to consist primarily of dark matter.
Here I summarise research that shows that initially spherical bound
low-mass satellites without dark matter, that are on orbits within a
massive Galactic dark corona, can evolve into remnants that are
non-spherical, have a non-isotropic velocity dispersion tensor and are
not in virial equilibrium, but are bright enough for sufficiently long
times to be mistaken for dark-matter dominated dSph galaxies.}

\vskip 6mm
\bigbreak
\noindent{\bf 1. Introduction}
\vskip 4mm
\noindent
At a distance ranging from a few tens to a few hundred~kpc the Galaxy
is surrounded by nine or more dSph galaxies which are barely
discernible stellar conglomerates, some appearing flattened and having
internal substructure. They lie in one or two planes and follow polar
orbits around the Milky Way.  The stellar content of these is
typically similar to that in globular clusters, apart from probably
much more complex star formation histories. However, the dSph galaxies
are roughly two orders of magnitude more extended (a few hundred pc)
and have about the same velocity dispersion ( 6--10~km/sec) as
globular clusters, perhaps implying large quantities of dark matter
(e.g. Mateo et al. 1993).  If this is true then this would be evidence
that dark matter can be a significant contributor to the system mass
over length scales of a few hundred~pc.

Recent reviews of dSph galaxies are published by Ferguson \& Binggeli
(1994) and Gallagher \& Wyse (1994).  Estimates of their structural
parameters and a compilation of kinematical data is provided by Irwin
\& Hatzidimitriou (1995).

One possible alternative to the dark matter hypothesis as an
explanation of the large mass-to-light ratia ($M/L$, hereafter always
in solar units) in dSph galaxies is that these stellar systems may not
be in virial equilibrium but instead are significantly perturbed by
Galactic tides.

Oh, Lin \& Aarseth (1995) model dSph galaxies orbiting in different
rigid spherical Galactic potentials. The dSph satellites are
represented with $10^3$~particles using a softened direct $N$-body
programme.  Oh et al. find that while the satellite remains a bound
stellar system it would not be mistaken as a dSph satellite with a
large observed (i.e. {\it apparent }) mass-to-light ratio $(M/L)_{\rm
obs}$.  Given the necessarily relatively small number of particles,
the very late stages of disruption and the subsequent evolution is
determined but by a few particles and cannot be followed in sufficient
detail because two-body relaxation becomes significant.  Piatek \&
Pryor (1995) simulate one perigalactic passage in different rigid
spherical Galactic potentials of a representative dSph galaxy, which
they model with $10^4$~particles using the TREECODE.  They find that a
single perigalactic passage cannot perturb a satellite such that an
observer would measure large apparent $(M/L)_{\rm obs}$ values.

Thus one would tend to conclude that Galactic tides cannot evolve
initially bound satellites with `normal' $M/L\approx 1-3$ into
dSph-like systems with $(M/L)_{\rm obs}>10$.  Nevertheless, the local
dSph galaxies show peculiar correlations between their location
relative to the Milky Way and their kinematical, structural and
photometric data that appear to bear the expected signature of
possibly significant tidal effects (Bellazzini, Fusi Pecci \& Ferraro
1996), and Burkert (1996) points out that most of the dSph satellites
that lie within a distance of 100~kpc appear to have tidal radii that
are too small to permit large quantities of dark matter.

In this summary I discuss the results of two simulations of a
representative low-mass satellite that is injected into a massive and
extended Galactic dark corona (often referred to as dark halo by other
authors) on different orbits. A full report of this work will be
available in Kroupa (1997).

\vskip 6mm
\bigbreak
\noindent{\bf 2. Method}
\vskip 4mm
\noindent
The conventional particle-mesh technique is well suited for the
simulation of the dynamical interaction of two galaxies because it is
fast and inherently collisionless so that two-body relaxation is
negligible. However, the resolution of the central regions of galaxies
is limited by the coarseness of the grid.  

The simulations described here are performed with a generalisation, by
R. Bien and later N. Wassmer, of the conventional particle-mesh
technique, making use of the additivity of the potential by employing
in total five sub-grids per galaxy. Three of these co-move with the
centre of density of a galaxy and the other two contain the local
universe.  The simulations here use $32^3$ grid cells in each active
grid. An early description of the code, called SUPERBOX, is provided
by Bien, Fuchs \& Wielen (1991, see also Madejski \& Bien 1993).

SUPERBOX computes the radii of the mass shells and the eigenvalues of
the moment of inertia and velocity dispersion tensors for each galaxy
in the simulation. Every pre-chosen number of integration steps a
subset of all particles of a galaxy are stored on disk.  These are
used to evaluate the projected radial surface brightness profile, the
line-of-sight and tangential velocity dispersions and the integrated
satellite brightness. Velocity dispersions are estimated using the
iterated bi-weight scale estimator (Beers, Flynn \& Gebhardt 1990).
The central surface brightness, which is obtained from the fitted
exponential radial brightness profile, the central line-of-sight
velocity dispersion and the half-brightness radius an observer sees
from Earth, are used to calculate the {\it apparent} mass-to-light
ratio, $(M/L)_{\rm obs}$, using the King-formula (equation~1 in Piatek
\& Pryor 1995), the application of which presupposes, among other
conditions, sphericity of the stellar system, an isotropic velocity
dispersion tensor and dynamical equilibrium. 

The Galactic dark corona is represented by a non-singular isothermal
density profile with circular speed $V_{\rm c}=220$~km/sec extending
to 250~kpc and using $10^6$ particles, and its nested grids have
dimensions of (50~kpc)$^3$ and (188~kpc)$^3$.  The initial satellite
is assumed to be a Plummer sphere, which is as good a description as
can reasonably be obtained for a dSph satellite, with a Plummer radius
$R_{\rm pl}=0.3\,$kpc, total mass $10^7\,M_\odot$ consisting of
$3\times10^5$ particles. Each particle is assumed to have a true
photometric V-band mass-to-light ratio $(M/L)_{\rm true}\le3$.  The
nested grids have dimensions of (1.6~kpc)$^3$ and (8~kpc)$^3$.  Both
the Galactic dark corona and the satellite are contained in an outer
grid with dimension (700~kpc)$^3$, and are individually allowed to
relax to equilibrium. An integration time-step of 1.1~Myr is
adopted. As a result of the mild contraction the Galactic dark corona
becomes slightly prolate. This is a reasonable basis for a model of
the dark matter distribution because cosmological structure formation
simulations indicate that dark coronae are likely to be non-spherical.

The long-axis of the dark corona is defined to be the x-axis.  The
satellite is injected 100~kpc from the centre of the dark corona along
the x-axis with a velocity vector parallel to the y-axis with a speed
of 100~km/sec and 125~km/sec, giving eccentricities~0.74 (simulation
`RS1-4') and~0.60 (simulation `RS1-5'), and maximum speed near
perigalacticon of about 480~km/sec (RS1-4) and 450~km/sec (RS1-5). The
orbital period is approximately 2.1~Gyr in both cases.

\vskip 6mm
\bigbreak
\noindent{\bf 4. Results}
\vskip 4mm
\noindent
Dynamical friction does not affect the orbit owing to the small
satellite mass. The satellite looses 10-20~per cent of its particles
each time it passes perigalacticon, and the time-varying Galactic tidal
field stimulates internal collective oscillations (compare with Kuhn
\& Miller 1989).  The central surface brightness increases temporarily during
each perigalactic passage with subsequent reduction due to mass loss,
the satellite re-adjusting its structure to new virial equilibrium.
During each perigalactic passage the velocity dispersions increase
temporarily (compare with Piatek \& Pryor 1995).  However, as already
stressed by Oh et al. (1995), the velocity dispersions show an overall
decrease as mass is lost and while the satellite remains bound and in
dynamical equilibrium.

After passage through perigalacticon at about 4.6~Gyr (RS1-4) and
6.8~Gyr (RS1-5) the satellite can be considered disrupted, although a
remnant remains (compare with Johnston, Spergel \& Hernquist 1995)
with half-light radius of a few hundred~pc.  The central surface
luminosity of the remnant decays by a factor of about~100 during the
following 3-4~Gyr, with an initial rapid decline that levels off. If
$(M/L)_{\rm true}<3$ then the remnants are bright enough for many
orbits to compare favourably with the observed dSph satellites.

After disruption near 4.6~Gyr (RS1-4) and 6.8~Gyr (RS1-5) the measured
central line-of-sight velocity dispersion of the remnant varies
significantly in the range of about 2 to 30~km/sec depending on
orbital eccentricity and phase. The velocity dispersion along the
orbit is significantly reduced compared to the velocity dispersion
along the line joining the Galactic centre and the satellite. This
means that the remnant can survive as a distinct density enhancement
for long periods (compare with Kuhn 1993). The remnant is
non-spherical but need not appear elongated along the orbital path.

Given the structure and line-of-sight kinematics of the remnant viewed
in the observational plane, an observer deduces apparent $(M/L)_{\rm
obs}>10$, with values as large as a few hundred, although $(M/L)_{\rm
true} \le 3$.

The plane of both satellite orbits tilts by about the same amount under
the action of the torques from the prolate Galactic corona and the
transfer of energy and angular momentum. The orbits become
increasingly polar but the two orbits remain approximately co-planar.

\vskip 6mm
\bigbreak
\noindent{\bf 5. Conclusions}
\vskip 4mm
\noindent
Fully self-consistent simulations of the interaction of dwarf
satellite galaxies without dark matter with an extended Galactic dark
corona indicate that the high $M/L$ values inferred for dSph
satellites do not necessarily imply that these contain dark
matter. Rather, high apparent $(M/L)_{\rm obs}$ values result
naturally from preferential orbital-phase modulated tidal removal of
satellite particles leading to highly non-uniform phase space
distribution of particles in the long-lived remnant.

The reason for large $(M/L)_{\rm obs}$ is that the model remnants have
non-isotropic velocity dispersions, are non-spherical and are not in
dynamical equilibrium, and may perhaps best be described as amorphous
lumps of stars.

The present model prolate Galactic dark corona leads to a roughly
planar distribution of remnants on near-polar orbits that bear
tantalizingly close resemblance to the observed dSph satellites.

\bigskip
\noindent{\bf REFERENCES}
\nobreak
\bigskip
\nex Beers T. C., Flynn K., Gebhardt K., 1990, AJ, 100, 32
\nex Bellazzini M., Fusi Pecci F., Ferraro F. R., 1996, MNRAS, 278, 947
\nex Bien R., Fuchs B., Wielen R., 1991, In: Tenner A. (ed.),
     Proc. CP90 Europhys. Conf. Comput. Phys., World Scientific,
     Singapore, p.3
\nex Burkert A., 1996, ApJL, in press
\nex Ferguson H. C., Binggeli B., 1994, A\&AR, 6, 67
\nex Gallagher III J. S., Wyse R. F. G., 1994, PASP, 106, 1225
\nex Irwin M., Hatzidimitriou D., 1995, MNRAS, 277, 1354 
\nex Johnston K. V., Spergel D. N., Hernquist L., 1995, ApJ, 451, 598
\nex Kroupa P., 1997, in preparation
\nex Kuhn J. R., 1993, ApJ, 409, L13
\nex Kuhn J. R., Miller R. H., 1989, ApJ, 341, L41
\nex Madejski R., Bien R., 1993, A\&A, 280, 383
\nex Mateo M., Olszewski E. W., Pryor C., Welch D. L., Fischer P.,
     1993, AJ, 105, 510 
\nex Oh K. S., Lin D. N. C., Aarseth S. J., 1995, ApJ, 442, 142
\nex Piatek S., Pryor C., 1995, AJ, 109, 1071

\vfill
\bye